\documentstyle[prl,aps,epsfig,multicol,ifthen,color]{revtex}

\input{epsf}
\begin{document}
\draft
\title{Experimental demonstration of ground state laser cooling with electromagnetically induced transparency}
\author{C.~F.~Roos\cite{byline}, D.~Leibfried, A.~Mundt, F.~Schmidt-Kaler, J.~Eschner, R.~Blatt}
\address{Institut f\"ur Experimentalphysik, University of Innsbruck, A-6020 Innsbruck, Austria}
\date{\today}
\maketitle
%
\begin{abstract}
Ground state laser cooling of a single trapped ion is achieved
using a technique which tailors the absorption profile for the
cooling laser by exploiting electromagnetically induced
transparency in the Zeeman structure of a dipole transition. This
new method is robust, easy to implement and proves particularly
useful for cooling several motional degrees of freedom
simultaneously, which is of great practical importance for the 
implementation of quantum logic schemes with trapped ions. 

\end{abstract}


\begin{multicols}{2}

\begin{center}
\begin{figure}[t]
\epsfig{file=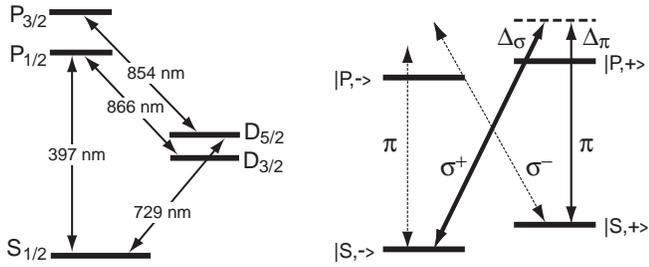,width=0.99\hsize}\vspace{\baselineskip}
\caption{\label{levels}Levels and transitions in $^{40}$Ca$^+$
used in the experiment (left). Zeeman sublevels of the S$_{1/2}$
and P$_{1/2}$ states and lasers relevant for the cooling (right).
The $\sigma^-$ light arises from the $\pi$ laser beam not being
orthogonal to the quantization axis. For EIT-cooling, 
$\Delta_{\pi} = \Delta_{\sigma}$.} 
\end{figure}
\end{center}

\begin{center}
\begin{figure}[t]
\epsfig{file=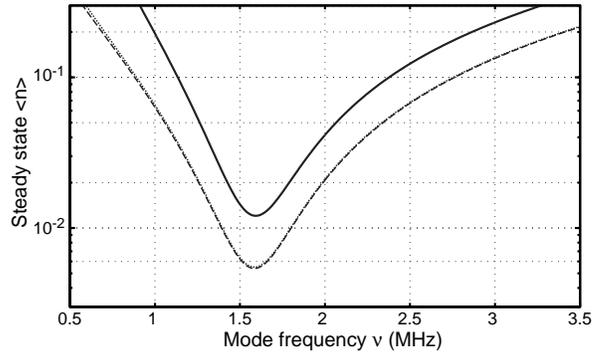,width=0.9\hsize}\vspace{\baselineskip}
\caption{\label{Theory}Steady state vibrational excitation
calculated for a 3-level system (dashed curve), a 4-level system
and ideal polarizations, i.e.\ $\pi$-light $\bot$ B-field
(dottet), and a 4-level system with $\pi$-light at $55^\circ$ to
B-field as in the experiment (solid). Parameters are
$\Delta_{\sigma} = \Delta_{\pi} = 2\pi \times 70$~MHz,
$\Omega_{\sigma} = 2\pi \times 21.4$~MHz, and $\Omega_{\pi} = 2\pi
\times 3$~MHz. The AC Stark shift is 1.6~MHz.}
\end{figure}
\end{center}

\begin{center}
\begin{figure}[t]
\epsfig{file=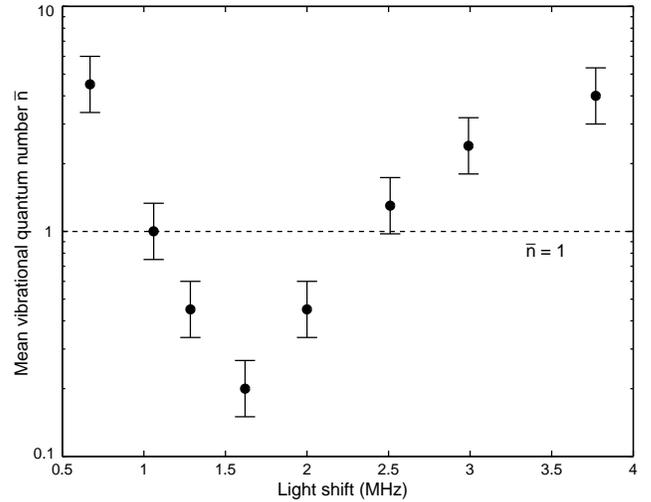,width=0.98\hsize}\vspace{\baselineskip} 
\caption{\label{deltany} Mean vibrational quantum number 
$\bar{n}_y$ vs. AC-Stark shift $\delta/2\pi$, after 7.9~ms of EIT 
cooling (starting from a thermal distribution with 
$\bar{n}_y=16$).} 
\end{figure}
\end{center}

\vspace{0mm} One of the most promising avenues towards
implementing the fundamental ingredients of a scalable quantum
computer is, as of today, based on trapped ions. With a 
string-like arrangement of several ions trapped in a 
radiofrequency (Paul) trap \cite{Paul}, deterministic entanglement 
between different ions \cite{DetermEntang} has been achieved. 
Individual addressing and state readout of ions in such a string 
with laser pulses has also been demonstrated \cite{Addressing}. 
Another fundamental requirement for quantum logic operations with 
trapped ions, following the original proposal \cite{Qcomputer}, is 
the preparation of the ions in the quantum mechanical ground state 
of their motion by laser cooling. While a single trapped ion was 
laser-cooled to the motional ground state as early as 1989 
\cite{Diedrich}, ground state cooling of an ion string 
\cite{Wineland2} and its combination with individual addressing 
\cite{Rohde} have only recently been demonstrated. The laser 
cooling methods used in those experiments are sideband cooling 
\cite{Diedrich,Rohde} and Raman sideband cooling \cite{Wineland2}, 
where a laser (or pair of lasers) exciting a narrow optical 
transition is detuned from the atomic resonance by the frequency 
of one motional quantum, thereby inducing transitions to 
lower-lying motional states until the ground state is reached. 
\\
\indent Although for quantum gate operation only one mode out of 
the $3N$ motional degrees of freedom of an $N$-ion string is 
required to be cooled to the ground state, high-fidelity 
manipulation of the qubits requires the other modes to be deep 
inside the so-called Lamb-Dicke regime, where their residual 
vibrational amplitude is very small compared to the wavelength of 
the laser that induces optical transitions \cite{spectator}. In 
conventional sideband and Raman sideband cooling, however, usually 
only one mode is cooled at a time, and the other modes are heated 
by spontaneous emission processes. A new cooling technique relying 
on electromagnetically induced transparency (EIT) 
\cite{EITcooling} eliminates these difficulties largely by 
providing a larger cooling bandwidth, such that several modes can 
be cooled simultaneously, and by suppressing, through quantum 
interference, a large fraction of the heating processes. In this 
Letter we describe the first experimental demonstration of this 
technique and show that apart from its advantageous properties 
regarding heating and bandwidth it is also technically 
significantly simpler, thus making it a very favourable cooling 
method for quantum logic experiments with single ions. 
\\
\indent The theoretical background of the method as described in 
\cite{EITcooling} has to be adapted only slightly to be applied to 
our experiment. We implemented the scheme on the ${\rm S}_{1/2}\to 
{\rm P}_{1/2}$ transition of a $^{40}$Ca$^+$ ion, whose Zeeman 
sublevels form a four-level system. We denote the levels by $|{\rm 
S},\pm\rangle$ and $|{\rm P},\pm\rangle$, see Fig.\ref{levels}. 
Three of the levels, $|{\rm S},\pm\rangle$ and $|{\rm 
P},+\rangle$, together with the $\sigma^+$- and $\pi$-polarized 
laser beams, form a system of the kind considered in 
\cite{EITcooling}, and the main modification is the fourth level 
whose effect will be discussed below. 
\\
\indent The principle of the cooling is, briefly, that the 
stronger blue-detuned $\sigma_+$ light (the coupling laser) 
creates a Fano-type absorption profile for the $\pi$ light (the 
cooling laser) which has a zero at $\Delta_{\pi} = 
\Delta_{\sigma}$ (this is the EIT condition) and a bright 
resonance corresponding to the dressed state 
$(\Omega_{\sigma}|{\rm S},-\rangle + 2\delta|{\rm 
P},+\rangle)/\sqrt{4\delta^2 + \Omega_{\sigma}^2}$ at 
$\Delta_{\pi} = \Delta_{\sigma}+\delta$, where $\delta = 
\frac{1}{2}(\sqrt{\Omega_{\sigma}^2 + \Delta_{\sigma}^2)} - 
|\Delta_{\sigma}|) \simeq \Omega_{\sigma}^2/4\Delta_{\sigma}$ is 
the AC Stark shift created by the $\sigma_+$ light. By choosing 
$\Omega_{\sigma} \simeq 2\sqrt{\omega\Delta_\sigma}$, the AC Stark 
shift $\delta$ is made equal to the vibrational frequency $\omega$ 
of the mode to be cooled. Then, with the cooling laser tuned to 
$\Delta_{\pi} = \Delta_{\sigma}$, no $\pi$ light is absorbed 
unless the ion provides a vibrational quantum, whereby the 
absorption probability is shifted to the bright resonance 
("sideband absorption") and the ion's motion is cooled 
\cite{linewidth}. The spontaneous emission completing an 
absorption-emission cycle happens predominantly without a change 
of the motional state if the Lamb-Dicke condition is fulfilled. 
Since absorption without a change in motional energy ("carrier" 
absorption) is cancelled by EIT, the heating which goes along with 
spontaneous emission after carrier absorption is also eliminated. 
This has consequences for both the mode to be cooled, for which a 
low final temperature can be reached, and for the other modes 
which are heated considerably less.\\ 
\indent It has been shown that for the cooling technique used here
\cite{EITcooling}, like for other methods \cite{Stenholm}, in the
Lamb-Dicke regime and below saturation of the cooling transition
the cooling process can be described by a rate equation of the
form
\begin{equation}
\label{Evolution} \frac{d}{dt}\langle n \rangle = - (A_- - A_+)
\langle n \rangle + A_+\;,
\end{equation}
where $\langle n(t) \rangle$ is the mean vibrational excitation of 
the mode under consideration. The coefficients $A_{\pm}$ are 
derived from a full quantum mechanical master equation 
\cite{EITcooling} and contain the quantum interference at 
$\Delta_{\pi} = \Delta_{\sigma}$. They can be interpreted as the 
rate coefficients for state-changing transitions induced by the 
cooling laser through sideband absorption and subsequent carrier 
emission \cite{Stenholm}. When the scattering rate (linewidth 
times population of the upper state) for the cooling laser is 
denoted by $W(\Delta_{\pi})$, we get for $\Delta_\pi = 
\Delta_\sigma$ 
\begin{equation}
\label{Aplusminus} A_{\pm} = \eta^2 \cos^2(\phi)
W(\Delta_{\pi}\mp\omega)\;,
\end{equation}
\noindent where $\eta=|\vec{k_g}-\vec{k_r}|a_0$ is the Lamb-Dicke 
parameter, with $a_0$ rms size of the ground state of the harmonic 
oscillator and $\vec{k_g}$ ($\vec{k_r}$) cooling (coupling) laser 
wave vector, and $\phi$ is the angle between $\Delta \vec{k} = 
\vec{k_g}-\vec{k_r}$ and the motional axis. 
\\
\indent The scattering rate $W(\Delta_{\pi})$ can be calculated
from optical Bloch equations, see for example \cite{Cohen} for the
three-level case. In Fig.{\ref{Theory}} we plot the steady state
vibrational excitation $\langle n(\infty) \rangle =\bar{n}$
calculated for our experimental system and for some idealized
cases. The reduced efficiency in our system is due to the angle of 
55$^\circ$ (rather than ideal 90$^\circ$) between the $\pi$ 
polarized laser beam and the magnetic field which creates a 
$\sigma^-$ component that excites $|{\rm S},+\rangle \to$ $|{\rm 
P},-\rangle$ transitions and leads to unwanted heating. In the 
calculations we have neglected the effect of the ${\rm P_{1/2}} 
\to {\rm D_{3/2}}$ transition because of the small branching ratio 
$({\rm P_{1/2}}\to{\rm D_{3/2}}) : ({\rm P_{1/2}}\to{\rm S_{1/2}}) 
= 1:16$.\\ 
\indent The ion trap used in our experiment is the 1.4~mm~sized
3D-quadrupole Paul trap, described in detail in \cite{Roos99}. For
the experiments described below, a single $^{40}$Ca$^+$ ion is
loaded into the trap, having oscillation frequencies ($\omega_x$,
$\omega_y$, $\omega_z$)/(2$\pi$) of (1.69, 1.62, 3.32) MHz. We
investigated EIT cooling of the $y$ and the $z$ oscillation. The
experiments proceed in three steps:
\\
\indent (i) We first Doppler precool the ion on the S$_{1/2}$ to 
P$_{1/2}$ transition at 397~nm (natural linewidth 20~MHz). The 
necessary UV light near 397~nm is generated by frequency doubling 
a Ti:Sapphire laser. This light is passed through an AOM driven at 
80 MHz to switch the light for the different steps of our 
experiment. The +1st diffraction order beam is then focussed and 
directed onto the ion. A detuning of approximately -20 MHz with 
respect to the ${\rm S}_{1/2} \to {\rm P}_{1/2}$ transition line 
is chosen for optimum Doppler cooling results. To avoid optical 
pumping into the D$_{3/2}$ states, we employ a grating stabilized 
diode laser near 866~nm \cite{ExpIonsIbk}. The Doppler cooling 
limit on this transition of 0.5~mK corresponds to mean vibrational 
quantum numbers of $\bar{n}_{z} \approx$ 3 and $\bar{n}_{x} 
\approx \bar{n}_{y} \approx$ 6. The cooling limits reached in our 
experiment are higher, due to the fact that the simple assumption 
of a two level system in the determination of the Doppler limit 
does not hold in our case. We experimentally determined the mean 
excitation numbers after Doppler cooling to be 
$\bar{n}_z=6.5(1.0)$ and $\bar{n}_y=16(2)$. 
\\
\indent (ii) After Doppler cooling we apply a bichromatic pulse of 
radiation around 397~nm for EIT  cooling that will be described in 
more detail below. 
\\
\indent (iii) Finally we analyze the motional state after EIT 
cooling by spectroscopy on the ${\rm S}_{1/2} \to {\rm D}_{5/2}$ 
quadrupole transition at 729~nm. There we can resolve the motional 
sideband structure and detect the final electronic state with an 
electron shelving technique \cite{Roos99}. The quadrupole 
transition is excited with 729~nm radiation from a frequency 
stabilized Ti:Sapphire laser with a bandwidth of $\delta \nu \leq$ 
100~Hz. A diode laser at 854 nm serves to repump the ion from the 
D$_{5/2}$ to the S$_{1/2}$ level. 
\\
\indent In more detail, step (ii) is technically realized in the 
following way: To generate the two blue detuned beams for EIT 
cooling we use the zero order beam out of the Doppler switch AOM, 
split it with a 50/50 beam splitter, pass the two emerging beams 
through another two AOMs at 86 MHz (cooling laser) and 92 MHz 
(coupling laser) respectively, and focus the +2nd diffraction 
orders onto the ion. The frequency difference equals the Zeeman 
splitting of the $|{\rm S},\pm \rangle$ levels in the quantization 
B-field (4.4 Gauss). By this arrangement, the Fano absorption 
profile for the cooling laser is created at approximately 75~MHz 
(about 3.5 natural linewidths) above the $|{\rm P}, + \rangle$ 
level. The $k$-vectors of the cooling beams enclose an angle of 
$125^\circ$ and illuminate the ion in such a way that their 
difference $\Delta \vec{k}$ has a component along all trap axes 
($(\phi_x,\phi_y,\phi_z) = (66^\circ,71^\circ,31^\circ)$, where 
$\phi_i$ denotes the angle between $\Delta \vec{k}$ and the 
respective trap axis). 
\\
\indent To roughly set the light intensity on the ion such that 
the AC-Stark shift of the $|{\rm P},+ \rangle$ level due to the 
$\sigma^+$ beam coincides with the trap frequency, we recorded 
spectra of the ${\rm S}_{1/2} \to {\rm D}_ {5/2}$ quadrupole 
transition with the dressing $\sigma^+$ laser beam switched on. 
The $\sigma^+$ beam shifts the $|{\rm S}, - \rangle$ level by an 
amount that is equal in magnitude to the AC Stark shift of the 
dressed $|{\rm P}, + \rangle$ level used in the cooling scheme. 
Therefore by determining the change in the carrier $|{\rm S},- 
\rangle \to {\rm D}_{5/2}(m=-5/2)$ transition frequency versus the 
$\sigma^+$ intensity we obtained a direct measure of the AC Stark 
shift relevant for EIT cooling \cite{nopumping}. 
\\
\indent Then we applied EIT-cooling to the radial $y$-mode of our 
trap with a frequency of $\omega_y = 2\pi\times 1.62$~MHz. After 
1.5~ms of Doppler cooling, we switched on both EIT cooling beams 
for 7.9~ms. The $\sigma^+$-beam was left on for 50~$\mu$s after 
the $\pi$-beam was turned off to optically pump the ion to the 
$|{\rm S},+\rangle$ ground state. We then monitored the final 
state by exciting Rabi-oscillations on the blue sideband of the 
$|{\rm S},+\rangle \to$ D$_{5/2}(m=+5/2)$ transition and measuring 
the $|{\rm S},+\rangle$ level occupation as a function of the 
pulse length \cite{Roos99}. The Rabi-oscillations were 
subsequently fitted to determine the mean vibrational occupation 
number $\bar{n}_y$ \cite{Meekhof}. We recorded $\bar{n}_y$ as a 
function of the AC-Stark shift $\delta$ by varying the intensity 
of the $\sigma^+$-beam around the value $\delta=\omega_y$ 
previously determined. Our results are depicted in Fig.\ 
\ref{deltany}, showing that the initial thermal distribution with 
$\bar{n}_y=16$ is cooled close to the ground state over a wide 
range of AC-Stark shifts. As expected the most efficient cooling 
occurs for $\delta \simeq \omega_y$. 
\\
\indent The lowest mean vibrational number $\bar{n}_y=0.18$ 
observed for $\delta=2\pi\times$~1.6~MHz corresponds to 84 \% 
ground state probability. We repeated this experiment on the 
3.3~MHz $z$-mode after having increased the intensity of the 
$\sigma^+$-beam. For this mode, a minimum mean vibrational number 
of $\bar{n}_z=0.1$ was obtained, corresponding to 90 \% ground 
state probability for $\delta=2\pi\times$~3.3~MHz. 

\begin{center}
\begin{figure}[t]
\epsfig{file=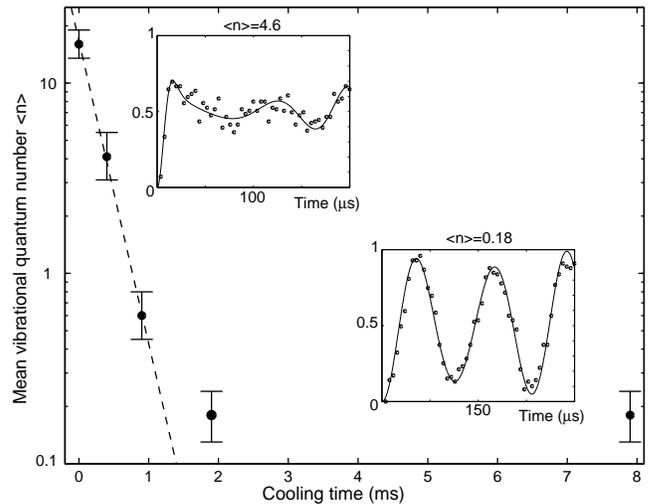,width=0.98\hsize}\vspace{\baselineskip} 
\caption{\label{coolingdynamics} Mean vibrational quantum number 
$\bar{n}_y$ vs. EIT cooling pulse length $\tau$. The insets show 
Rabi oscillations excited on the upper motional sideband of the 
$|{\rm S},+\rangle \to$ D$_{5/2}(m=+5/2)$ transition, after 0.4 ms 
(left) and 7.9 ms (right) of EIT cooling. A thermal distribution 
is fitted to the data to determine $\bar{n}_y$. } 
\end{figure}
\end{center}

\indent The cooling results are largely independent of the
intensity of the $\pi$-beam as long as it is much smaller than the
$\sigma^+$ intensity. In our experiment the intensity ratio was
$I_{\sigma}/I_\pi \simeq 100$ and we varied the intensity of the
$\pi$-beam by a factor of 4, with no observable effect on the
final $\bar{n}$.
\\
\indent By determining the dependence of the mean vibrational 
quantum number on the EIT cooling pulse length $\tau$, we measured 
the cooling time constant to be 250 $\mu$s, see Fig. 
\ref{coolingdynamics}.
\indent In order to show that the EIT method is suitable to 
simultaneously cool several vibrational modes with vastly 
different frequencies of oscillation, we chose the axial $z$-mode 
at 3.3 MHz, and the radial $y$-mode at 1.62 MHz, which have a 
frequency difference of 1.7 MHz. The intensity of the 
$\sigma^+$-beam was set such that the AC Stark shift was 2.6 MHz, 
roughly halfway between the two mode frequencies. Again we applied 
the EIT cooling beams for 7.9 ms after Doppler cooling. This time 
we determined the final $\bar{n}$ by comparing the excitation 
probability on the red and the blue sideband of the ${\rm 
S}_{1/2}(m=1/2) \to {\rm D}_{5/2}(m=5/2)$ transition 
\cite{Diedrich}. We find both modes cooled deeply inside the 
Lamb-Dicke regime ($\eta \sqrt{\bar{n}} \ll 1$), with 
$p_0^{(y)}=58\%$ and $p_0^{(z)}=74\%$ ground state probability 
\cite{xmode}. 
\\
\indent In conclusion we have experimentally demonstrated a novel 
cooling method for trapped particles which allowed us to 
efficiently cool a trapped ion to the ground state of motion, 
using only a Zeeman-split dipole transition. We also showed that 
the cooling mechanism has a considerable bandwidth and should thus 
allow to simultaneously cool several modes of longitudinal ion 
motion in a linear trap. This is important for implementing 
quantum logical gates with trapped ions, because the requirement 
of individual optical addressing puts an upper limit on the 
motional frequencies that usually makes it desirable to cool these 
modes below the Doppler limit. The cooling method does not require 
a forbidden transition and involves a lower technical overhead as 
compared to other ground state cooling methods demonstrated so 
far. Furthermore, the demonstrated method is not restricted to 
trapped ions and should work for trapped neutral particles as 
well. Its large bandwidth should make this scheme especially 
attractive for optical lattices since the slight anharmonicity and 
site-to-site inhomogeneity of the lattice potentials does not 
hinder effective cooling.   
\\
\indent This work is supported by the Austrian 'Fonds zur 
F\"orderung der wissenschaftlichen Forschung' (SFB15 and 
START-grant Y147-PHY), by the European Commission (TMR networks 
'Quantum Information' (ERB-FRMX-CT96-0087) and 'Quantum 
Structures' (ERB-FMRX-CT96-0077)), and by the "Institut f\"ur 
Quanteninformation GmbH".


%
\end{multicols}

\begin{references}

\bibitem[*]{byline} Present address: Laboratoire Kastler-Brossel,
Ecole Normale Sup\'erieure, 24 rue Lhomond, 75005 Paris, France

\bibitem{Paul} W. Paul, Rev. Mod. Phys. {\bf 62}, 531 (1990).

\bibitem{DetermEntang}
C. A. Sackett {\it et al.}, Nature {\bf 404}, 256 (2000).

\bibitem{Addressing}
H. C. N{\"a}gerl {\it et al.}, Phys. Rev. A {\bf 60}, 145 (1999).

\bibitem{Qcomputer}
J. I. Cirac and P. Zoller, Phys. Rev. Lett. {\bf 74}, 4091 (1995).

\bibitem{Diedrich}
F. Diedrich, J. C. Bergquist, W. M. Itano, and D. J. Wineland,
Phys. Rev. Lett. {\bf 62}, 403 (1989).

\bibitem{Wineland2}
Q. A. Turchette {\it et al.}, Phy. Rev. Lett. {\bf 81}, 1525
(1998).

\bibitem{Rohde}
F. Schmidt-Kaler {\it et al.}, quant-ph/0003096.

\bibitem{spectator}
D. J. Wineland {\it et al.}, J. Res. Natl. Inst. Stand. Technol.
{\bf 103}, 259 (1998).

\bibitem{EITcooling}
G. Morigi, J. Eschner and C. Keitel, quant-ph/0005009.

\bibitem{linewidth}
By adjusting the detuning $\Delta_\sigma$, the linewidth 
$\Gamma^\prime \simeq \Gamma(\frac{2\delta}{\Omega_\sigma})^2 
\simeq \Gamma\omega/\Delta_\sigma$ of the dressed state can be 
varied. 

\bibitem{Stenholm}
S. Stenholm, Rev. Mod. Phys. {\bf 58}, 699 (1986).

\bibitem{Cohen}
B. Lounis and C. Cohen-Tannoudji, J. Phys. II (France) {\bf 2},
579 (1992).

\bibitem{Roos99}
Ch. Roos {\it et al.}, Phys. Rev. Lett. {\bf 83},  4713 (1999).

\bibitem{ExpIonsIbk}
H. C. N\"agerl {\it et al.}, Appl. Phys. B {\bf 66}, 603 (1998).

\bibitem{nopumping}
Optical pumping to the $|{\rm S},+\rangle$ level was negligible 
during the 10 $\mu$s excitation pulse. 

\bibitem{Meekhof}
D. M. Meekhof {\it et al.}, Phys. Rev. Lett. {\bf 76}, 1796 
(1996). 

\bibitem{xmode}
Though we did not investigate the motional state of the 
x-oscillator (which was orthogonal to the beam used for the 
analysis), we expect the mode to be cooled to a similar degree as 
the y-oscillator. 

\end{references}
\end{document}